\documentclass[12pt]{article}
\usepackage{a4wide}
\usepackage{amssymb}
\usepackage{cite}
\begin{document}
{\renewcommand{\thefootnote}{\fnsymbol{footnote}}
\begin{center}
{\LARGE  Loop quantum gravity, signature change,\\ and the no-boundary proposal}\\
\vspace{1.5em}
Martin Bojowald$^1$\footnote{e-mail address: {\tt bojowald@gravity.psu.edu}}
 and 
Suddhasattwa Brahma$^2$\footnote{e-mail address: {\tt
    suddhasattwa.brahma@gmail.com}}\\
\vspace{0.5em}
$^1$ Institute for Gravitation and the Cosmos,
The Pennsylvania State
University,\\
104 Davey Lab, University Park, PA 16802, USA\\
\vspace{0.5em}
$^2$ Department of Physics, McGill University, Montr\'{e}al, QC H3A 2T8, Canada
\vspace{1.5em}
\end{center}
}

\setcounter{footnote}{0}

\begin{abstract}
  Covariant models of loop quantum gravity generically imply dynamical
  signature change at high density. This article presents detailed derivations
  that show the fruitful interplay of this new kind of signature change with
  wave-function proposals of quantum cosmology, such as the no-boundary and
  tunneling proposals. In particular, instabilities of inhomogeneous
  perturbations found in a Lorentzian path-integral treatment are naturally
  cured. Importantly, dynamical signature change does not require Planckian
  densities when off-shell instantons are relevant.
\end{abstract}

\section{Introduction}

The no-boundary proposal \cite{nobound} shares with models of loop quantum
gravity \cite{Action,SigChange,SigImpl} a prominent role played by signature
change. It is therefore of interest to discuss crucial technical differences
but also unexpected synergies between these two approaches.

A general difference is given by how signature change features in these
approaches to quantum cosmology --- as a postulate in the no-boundary proposal
and as an unexpected, derived result in models of loop quantum cosmology.  The
no-boundary proposal, in its simplest form with only a cosmological constant
as the energy ingredient, implements specific initial conditions for the wave
function of a universe by closing off an expanding Lorentzian space-time with
a compact Euclidean cap at the big bang. Euclidean signature is introduced
because it makes it possible to have 4-sphere-like solutions that can provide
such a closing cap. Here, signature change happens in a discontinuous manner
by gluing together two semiclassical geometries with different signatures.

Models of loop quantum cosmology, by contrast, do not include Euclidean
signature in their set-up but rather derive, after what usually amounts to a
long analysis, a modified space-time structure from other quantum effects in
the dynamical equations that define such models. As one generic outcome,
models of loop quantum cosmology may exhibit space-time structures with
non-singular signature change mediated by a curvature-dependent
function that crosses zero in a continuous manner. For the sake of clarity, we
will refer to the version of signature change realized in models of loop
quantum gravity as {\em dynamical signature change}.

Another difference is that the no-boundary proposal contains signature change
in the form of imaginary time, or a Wick rotation used in a Euclidean path
integral, while models of loop quantum gravity work with real proper time.
The reference to imaginary time in the no-boundary proposal can be eliminated
by formulating it in a Lorentzian, as opposed to Euclidean, path integral
\cite{LorentzianQC,RealNoBound}, suggesting a potentially more direct
correspondence with models of loop quantum gravity. However, in this case the
no-boundary proposal is defined for off-shell instantons that solve the
Raychaudhuri equation but not the Friedmann equation in order to implement the
no-boundary initial condition of zero scale factor. In loop quantum cosmology,
by contrast, the usual derivation of signature change requires certain
identities that follow from a modified Friedmann equation (or its extension to
inhomogeneous modes) and seem unavailable if only the Raychaudhuri equation
is used.

On a more physical level, the no-boundary proposal works at sub-Planckian
densities, assuming a small cosmological constant as the main source of
stress-energy of a young universe. This is in fact one of its appealing
features because it makes the proposal insensitive to detailed properties of
quantum gravity. In models of loop quantum gravity, by contrast, signature
change as discussed so far happens very close to Planckian densities. It
relies on certain effects of loop quantum gravity not realized in classical
general relativity, where signature change would be singular
\cite{ClassSigChange,GeometrySig,BoundarySigChange,ClassSigChange2}.

Given these discrepancies, an application of loop quantum gravity to the
no-boundary proposal might seem useless. However, the relationship turns out
to be closer than it appears at first sight, and it is actually fruitful: An
application of the Lorentzian path integral to formulate the no-boundary
proposal eliminates imaginary time giving the initial state of the universe a
more physical interpretation. But a new problem then arises because
inhomogeneous perturbations around no-boundary instantons have been found to
be unstable \cite{NoSmooth,NoRescue,Tucci_Lehners} which implies, even in the
best-case scenario, that additional modifications to the initial conditions
are needed to avoid this instability at late-times \cite{RealNoBound,
  Vilenkin1, Vilenkin2}. This result endangers the proposal as a well-defined
initial scenario of the universe.

Unexpectedly, however, several properties of dynamical signature change in
models of loop quantum cosmology conspire to solve the stability problem of
the no-boundary proposal \cite{LoopsRescue}.  In the present paper, we
highlight the details of several new properties of signature change for
off-shell instantons in models of loop quantum gravity that are relevant for
this conclusion. Importantly, this new version of dynamical signature change
does not require Planckian densities, even though it is derived from the same
loop-inspired modifications that lead to Planckian dynamical signature change
in models \textit{not} based on the no-boundary condition. On a technical
level, our main results are therefore that (i) dynamical signature change can
be derived for off-shell instantons in models of loop quantum gravity and (ii)
can have markedly different features in this setting compared with on-shell
solutions of modified constraint equations.  We expand on our previous work
\cite{LoopsRescue} to give details of the saddle-point analysis and the
structure of our new off-shell instantons. Moreover, we generalize our system
to include and parameterize different quantization ambiguities associated with
loop quantum gravity. We also point out several subtleties in the detailed
derivation given here, which may be technical but are nevertheless surprising
and essential for our physical results.

\section{Instabilities in the Lorentzian path integral}

We begin with a brief review of the Lorentzian path integral, following
\cite{LorentzianQC,RealNoBound}. Boundary conditions for the no-boundary
proposal can be formulated in a Lorentzian path integral of the usual
integrand $\exp(iS/\hbar)$, as opposed to the Euclidean path integral of
$\exp(-S/\hbar)$. Such an integral of a phase factor over real configurations
usually converges very slowly, but convergence can be improved by an
application of Picard--Lefshetz theory, shifting the integration contour onto
the complex plane. Because the value of the integral is not changed thanks to
Cauchy's theorem, provided the action $S$ does not introduce poles in the
complex plane, it remains Lorentzian even though complex configuration
variables appear in the improved integration. It is therefore possible to
avoid the use of imaginary time, and the original Lorentzian value of the path
integral remains unmodified.

Preparing for an application of Picard--Lefshetz theory, the Lorentzian
formulation of the no-boundary proposal deals with the problem of time by
fixing time reparameterization invariance almost completely, specifying that
the lapse function equals $N(t)=M/a(t)$ with a positive constant (rather than
time-dependent function) $M$. (The factor of $1/a$ is introduced for
convenience, and for the same reason transformed to the new variable $q=a^2$
\cite{NoSmooth,NoRescue}.)  Only this {\em constant} $M$, rather than a
time-dependent function $N(t)$, is integrated over in path integrals. Because
there is no longer a free multiplier at any given time, integration over $M$
does not impose the Hamiltonian constraint or the Friedmann equation except at
one time, which can be chosen to be the final moment included in a given path
integral. This is the reason why path integrals in this formulation describe
off-shell instantons, as a consequence of a specific approach to the problem
of time.  

A no-boundary instanton is then defined by the initial condition
$q(0)=0$, together with some fixed value of $q(1)=q_1$. For small $t$, a
linear function $q(t)=q_1t$ can be seen to solve the Raychaudhuri equation
\begin{equation} 
 \frac{{\rm d}^2q}{{\rm d}t^2}=2\Lambda 
\end{equation}
for $q=a^2$ in the presence of a cosmological constant $\Lambda$ (as well as
positive spatial curvature). However, this initial condition cannot be
compatible with the Friedmann equation for the same ingredients, given by
\begin{equation} \label{Friedmann1}
 \left(\frac{{\rm d}q}{{\rm d}t}\right)^2=4(\Lambda q- 1)\,.
\end{equation}
The instantons considered in the Lorentzian formulation therefore must be
strictly off-shell. (In the original Euclidean formulation, the condition
$q=0$ is compatible with the Wick-rotated version of (\ref{Friedmann1}).)

If the initial state is normalized, for simplicity assumed to be Gaussian,
repeated path integrations over real configuration variables should preserve
the Gaussian form. No instabilities could then arise.  However, as it turns
out, the equations of motion of inhomogeneous perturbations have solutions
with branch cuts on the positive-$M$ axis \cite{NoRescue}.  Picard--Lefshetz
theory determines that this branch cut should be circumvented through the
upper imaginary half-plane in the complex $M$-space. Along this contour, the
action has a negative imaginary part, and the bounded $\exp(iS/\hbar)$ is
turned into an unbounded upside-down Gaussian. Instabilities are then
inevitable, a conclusion which refers to general properties of the space-time
structure and also applies to alternative proposals of initial conditions,
such as the tunneling proposal \cite{tunneling}.

In more detail, the tensor mode equation for $h$ in the time variable defined
by choosing a lapse function $N=M/\sqrt{q}$,
written conveniently for $v=qh$, is
\begin{equation} \label{vclass}
 \ddot{v}-\frac{\ddot{q}}{q} v-\frac{M^2}{q^2} \beta \nabla^2v=0
\end{equation}
where we include a parameter $\beta=\pm1$ to see possible implications of
space-time signature. For Lorentzian signature, $\beta=1$, while $\beta=-1$
for Euclidean signature. In our subsequent application of methods of loop
quantum cosmology, $\beta$ will be a continuous function with $\beta\to1$ at
small curvature (late times).

For small $t$, the mode equation implies that
\begin{equation} \label{vclassapprox}
\ddot{v}_{\ell}\approx  -\frac{\beta\ell(\ell+2)M^2}{q_1^2} 
\frac{v_{\ell}}{t^2}
\end{equation}
for a fixed multipole number $\ell$, solved by any superposition of the two
independent solutions $v_{\ell,\pm} = t^{\frac{1}{2}(1\pm\gamma_{\ell})} v_1$ where
$v_1=v(1)$ with
\begin{equation} \label{gamma1}
 \gamma_{\ell}=\sqrt{1-4\beta\frac{\ell(\ell+2)M^2}{q_1^2}}\,.
\end{equation}
The action for (\ref{vclassapprox}), derived from tensor modes restricted to
leading terms at small $t$ is
\begin{equation}
 S_{\ell}=\frac{1}{16\pi G} \int_0^1 \left(\frac{\dot{v}_{\ell}^2}{M}-
   \beta\ell(\ell+2)M\frac{v_{\ell}^2}{q^2}\right)
= \frac{1}{16\pi GM} \int_0^1 \left(\dot{v}_{\ell}^2+
   \frac{\gamma_{\ell}^2-1}{4t^2}v_{\ell}^2\right)\,,
\end{equation}
inserting a small-$t$ off-shell instanton with $q(t)=q_1t$ in the second step.
The appearance of $M$ follows the characteristic dependence of matter or
perturbation actions on the lapse function of a background metric.  

Evaluated in $v_{\ell,\pm}$, we have
\begin{equation} \label{S}
 S_{\ell,\pm}= \frac{1}{32\pi GM}
 (1\pm\gamma_{\ell})\left.t^{\pm\gamma_{\ell}}\right|_{t=0}^1  v_1^2\,.  
\end{equation}
Only $v_{\ell,+}$ leads to a finite action $S_{\ell,+}$ because
$\gamma_{\ell}>0$; we therefore discard $v_{\ell,-}$. For complex $M$ in the
upper half plane, as dictated by Picard--Lefshetz theory,
\begin{equation}
 S_{\ell,+}\propto \frac{1}{M}= \frac{{\rm Re}M-i{\rm Im}M}{|M|^2}
\end{equation}
has a negative imaginary part which implies an unbounded $\exp(iS/\hbar)$.

Through $\gamma_{\ell}$, the solutions $v_{\ell,\pm}$ have a branch cut at
positive $M$ for $\beta>0$, in particular for $\beta=1$ as used in the
Lorentzian path-integral version of the no-boundary proposal. However, we can
already see that any dynamical signature change that would turn $\beta$ to
negative values at small $t$ can resolve the problem: If this happens, there
is no branch cut on the real $M$-axis.  An imaginary action and the
corresponding instability could then be avoided. However, unlike in models of
loop quantum gravity used so far, this dynamical signature change should
happen (i) for off-shell instantons and (ii) at sub-Planckian curvature
determined by a small cosmological constant. Each of these two conditions
requires a detailed analysis.

\section{Off-shell instantons in models of loop quantum gravity}

In models of loop quantum gravity, the Hamiltonian constraint is modified by
different effects motivated by mathematical properties of discrete space, most
importantly inverse-triad corrections \cite{InvScale} and holonomy
modifications \cite{GenericBounce}.  Holonomy modifications are used to
describe implications of the fact that loop quantum gravity implements
operators not for the gravitational connection $A_a^i$ (or extrinsic curvature
$K_a^i$) but only for its SU(2)-holonomies in space
\cite{LoopRep,ALMMT}. Holonomies are based on parallel transport,
\begin{equation} \label{Hol}
 h_e={\cal P}\exp\int_e A_a^i \tau_i \dot{e}^a{\rm d}\lambda\,,
\end{equation}
and are therefore integrated over spatial curves $e$ (with path ordering
${\cal P}$ of the non-commuting ${\rm su}(2)$ generators $\tau_i$). As
functions of the gravitational connection $A_a^i$, they are non-linear and
non-local.

In a homogeneous cosmological model, non-locality in space is not visible
because a position-independent connection then appears just as a collection of
spatial constants in the exponent of (\ref{Hol}). If the model is isotropic,
for instance, any connection with positive spatial curvature can be expressed
as $A_a^i=(\frac{1}{2}+c)\delta_a^i$ with a single canonical degree of freedom
$c=\dot{a}$ \cite{IsoCosmo}, using a basis adapted to the symmetry. This
degree of freedom is closely related to extrinsic curvature,
$K_a^i=c\delta_a^i$, which is more convenient in setting up a class of states
used in what follows \cite{FundamentalDisc,Spin}. Up to a constant phase
factor, holonomies along integral curves of the symmetry generators, taken in
an irreducible spin-$j$ matrix representation of SU(2), then have matrix
elements of the form $M_e=\exp(i j \ell_0\mu' c)$ with real numbers
$-1\leq\mu'\leq 1$. The fixed parameter $\ell_0$ determines the coordinate
size of a region in space which is taken as representative of the entire
homogeneous geometry. (This region may but need not be the entire
space. Nevertheless, to be specific, it can be taken to equal
$\ell_0=\sqrt[3]{2\pi^2}$ in the closed model.) The variable $\mu'$ then
determines the length of the curve as a fraction of the reference length
$\ell_0$. Combining $\mu'$ with $j$, any real number $\mu=j\mu'$ can be
achieved as a coefficient in the exponent of
\begin{equation} \label{HolIso}
 M_{\mu}=\exp(i\ell_0\mu c)\,.
\end{equation}

\subsection{Isotropic loop quantum cosmology}

In a reduction of the symplectic structure of general relativity to
isotropic models, $c$ is seen to be canonically conjugate to a momentum
$p\propto a^2$, such that
\begin{equation} \label{Poisson}
 \{c,p\}= \frac{8\pi G}{3\ell_0^3}\,.
\end{equation}
Alternatively, because $\ell_0c$ automatically appears in matrix elements of
holonomies, we may view $\ell_0c$ and $\ell_0^2p$ as canonical variables
with a Poisson bracket independent of $\ell_0$. 

\subsubsection{Representation}

We use the Poisson structure to set up a canonical quantization
\cite{IsoCosmo,Bohr}, modeling properties of holonomies as operators in loop
quantum gravity \cite{LoopRep,ALMMT}. In this theory, a holonomy function
(\ref{HolIso}) represents a normalizable state, unlike what would usually be
the case in the standard representation of a canonical bracket
(\ref{Poisson}). Loop quantization therefore exploits the existence of an
inequivalent representation of (\ref{Poisson}) that exists if one does not
require a quantization of (\ref{HolIso}) as a basic operator to be continuous
in $\mu$; see for instance \cite{BohrQM}.

This representation can be constructed by acting on the non-separable Hilbert
space of almost-periodic functions of $c$, defined as the space of functions
linearly generated by the basis (\ref{HolIso}) for all real $\mu$ with inner
product
\begin{equation}
 \langle M_{\mu_1},M_{\mu_2}\rangle= \lim_{C\to\infty} \frac{1}{2C}
 \int_{-C}^C M_{\mu_1}^*(c)M_{\mu_2}(c){\rm d}c\,.
\end{equation}
The momentum $\ell_0^2p$ acts on the basis states via
\begin{equation} \label{pspec}
 \ell_0^2\hat{p} M_{\mu}= \frac{8\pi G}{3} \frac{\hbar}{i} \frac{{\rm d}}{{\rm
     d}(\ell_0c)} \exp(i\ell_0\mu c)= \frac{8\pi}{3}\ell_{\rm P}^2 \mu
 M_{\mu}=\ell_0^2p_{\mu}M_{\mu}\,.
\end{equation}
The spectrum of $\ell_0^2\hat{p}/\ell_{\rm P}^2$ therefore contains all real
numbers
\begin{equation}
 \frac{p_{\mu}}{\ell_{\rm P}^2}= \frac{8\pi}{3}\mu\,,
\end{equation}
but it is discrete because the eigenstates $M_{\mu}$ are normalizable. (These
two properties are compatible with each other because the Hilbert space is
non-separable.)

To summarize, the non-linearity of holonomies is implemented through the use
of almost-periodic functions as states, on which the action of
$\hat{M}_{\mu}$,
\begin{equation}
 \hat{M}_{\mu_1} M_{\mu_2}= M_{\mu_1+\mu_2}\,,
\end{equation}
is not weakly continuous at $\mu_1=0$. (In the Hilbert space used here, any
two basis states, $M_{\mu_2}$ and $M_{\mu_1+\mu_2}$, are orthogonal for any
$\mu_1\not=0$.) Therefore, it is not possible to derive a quantization of the
linear phase-space function $c$ by taking a derivative of $M_{\mu}$ at
$\mu=0$. Only the non-linear functions $M_{\mu}(c)$ are represented as basic
operators in addition to $\hat{p}$, which have to be used to construct
possible quantizations of the polynomial Hamiltonian constraint (the Friedmann
equation) of a cosmological model.  

By replacing any polynomial reference to
$c=\dot{a}$ in a classical expression by periodic functions, holonomy
modifications are introduced.  There is much freedom in choosing a periodic
function to replace a polynomial, with the only condition that the classical
polynomial should be obtained as an approximation for small $c$ (or some other
function of $c$ being small, representing curvature).  A common choice is to
replace $c$ in the Friedmann equation with $\sin(\ell(a)c)/\ell(a)$, where
$\ell(a)$ is interpreted as a function that describes quantization ambiguities
as well as properties of an underlying discrete state. In particular, the
common choice $\ell(a)\approx \ell_{\rm P}/a$ with the Planck length
$\ell_{\rm P}$ \cite{APSII} leads to corrections in the Friedmann equation
that can be expanded in the dimensionless product $\ell_{\rm P}^2H^2$ with the
Hubble parameter $H$. Such corrections are relevant only near Planckian
curvature.

\subsubsection{Inverse-$a$ corrections}

The momentum operator $\hat{p}$ is represented directly as a basic
operator. Moreover, it has all real numbers as eigenvalues, such that one
could expect discretization effects to be minimal even though the spectrum is
formally discrete. However, the discreteness can lead to significant quantum
corrections whenever an inverse of $p$ is quantized. Because $\hat{p}$ has a
discrete spectrum containing zero, it does not have a densely defined inverse
operator. Nevertheless, we need to quantize inverse powers of $a$ or $p$ that
appear in the Hamiltonian constraint, for instance in the matter energy.

Using methods of \cite{QSDV}, it is possible to construct densely-defined
operators such that their classical limits reproduce the required inverse
power of $a$. The construction of these operators exploits the existence of
commutator identities such as
\begin{equation} \label{Inverse}
 \hat{M}_{\mu}^{-1}[\hat{M}_{\mu},\ell_0\sqrt{\hat{p}}]=
 -\frac{4\pi}{3\ell_0}\ell_{\rm P}^2 \mu
\widehat{p^{-1/2}}\,.
\end{equation}
On the left-hand side, all operators are densely defined on the Hilbert space
(taking the square root of $\hat{p}$ through the spectrum), and according to
the right-hand side their combination quantizes an expression with classical
limit proportional to $1/a$. An explicit calculation shows that the classical
limit is well approximated when the commutator acts on a state $M_{\mu_2}$
with $\mu_2\gg\mu$.  For small $\mu_2<\mu$, the behavior deviates from the
classical limit, implying inverse-$a$ corrections as a consequence of discrete
spatial geometry \cite{InvScale}. 

As with holonomy modifications, the representation of a given inverse power of
$a$ as an operator via commutator identities is not unique, leading to
additional quantization ambiguities. A detailed analysis of eigenvalues shows
that inverse-$a$ corrections can be parameterized broadly by a function
$f(a)$, such that \cite{Ambig,ICGC}
\begin{equation}
 (a^{-1})_{\mu} = \frac{1}{a_{\mu}} f(a_{\mu})
\end{equation}
where $a_{\mu}$ is obtained by taking the square root of a
$\hat{p}$-eigenvalue $p_{\mu}$ in (\ref{pspec}).  For large eigenvalues,
$f(a_{\mu})\sim1$, while the small-$\mu$ behavor is a power law
$f(a_{\mu})\approx a_{\mu}^{2n}$ with a positive integer $n>1$.

Holonomy modifications and inverse-$a$ corrections result from modelling
quantum-geometry effects of loop quantum gravity in isotropic situations. As
always in an interacting theory, quantum corrections also arise from quantum
back-reaction of fluctuations and higher moments of a state on the expectation
values \cite{EffAc,Karpacz}. Here, in accordance with the idea that the
no-boundary proposal can describe the origin of space-time without strong
quantum effects, we are interested in regimes in which quantum back-reaction
is sub-dominant compared with geometrical effects.  

We will therefore analyze modified Friedmann equations in which inverse-$a$
corrections have been inserted, mainly in the curvature and matter terms, and
holonomy modifications have been used.  For the latter, to be specific we will
work with a function $\sin(\ell(a)c)/\ell(a)$ where the $a$-dependence of
$\ell(a)$ is of power-law form, $\ell(a)=\ell_0\delta (\ell_0a)^{2x}$ with
constants $\delta$ (scaling like $\ell_0^{-1-2x}$) and $x$. The exponent $x$
is a parameter in an effective description and may therefore be running, such
that it may take different values in different ranges of a curvature scale
\cite{EFTLQC,Claims}. We will be able to ignore the running because our main
results apply asymptotically close to the initial state of the no-boundary
proposal. We will, however, take into account the possibility of having
different constant values of $x$, depending on the theory and the underlying
quantum-gravity state.  As we will review in more detail later on, when the
distinction will become important, two common choices for $x$ are $x=0$ (a
constant co-moving length $\ell(a)$) and $x=-1/2$ (a constant geometrical
length $a\ell(a)$).

\subsubsection{Modified background dynamics}

If $\ell$ depends on $a$ in power-law form, it is convenient to introduce
canonical variables such that $\ell(a)c$ is proportional to the new momentum:
\begin{equation} \label{QP}
 Q=\frac{3(\ell_0a)^{2(1-x)}}{8\pi G(1-x)} \quad\mbox{and}\quad
 P=-\ell_0^{2x+1}a^{2x}\dot{a}\,. 
\end{equation}
Moreover, we will use the definition $\bar{Q}=\frac{8}{3}\pi
G(1-x)Q=(\ell_0a)^{2(1-x)}$ in order to obtain more compact equations.  In
these variables, the classical constraint
\begin{equation} \label{ConsClass} 
C_{\rm class}=\ell_0^3\left(-\frac{3}{8\pi
      G} a\left(\dot{a}^2+k\right)+ a^3 \rho\right) =0\,,
\end{equation}
is modified to
\begin{equation} \label{Cons}
 C=-\frac{3}{8\pi G} \left(\bar{Q}^{(1-4x)/(2(1-x))} \frac{\sin^2(\delta
     P)}{\delta^2}+\bar{Q}^{1/(2(1-x))} \kappa(Q)\right)+ m(Q)g(Q)=0
\end{equation}
with inverse-$a$ corrections in the curvature term, $\kappa(Q)$, and in the
matter term, $g(Q)$ multiplying the matter energy $m=\ell_0^3a^3\rho$
contained in the averaging region of a homogeneous model. Classically,
$g(Q)=1$ while $\kappa(Q)=\ell_0^2$ for positive spatial curvature. In
(\ref{ConsClass}), we have included a factor of the coordinate volume
$\ell_0^3$ since the Hamiltonian constraint is spatially integrated.

Because the curvature term and the matter term depend on $a$ through different
powers, $\kappa(Q)\not=g(Q)$ in general. It is also possible that the
$P$-dependent term in (\ref{Cons}) is modified not just by holonomies
(non-zero $\delta$) but also by inverse-$a$ corrections. Such a term may be
required if there are explicit inverse powers of $Q$ in (\ref{Cons}), or if
one extends the isotropic models used here to anisotropic ones, in which case
even the classical constraint will have additional inverse powers of the
anisotropic scale factors \cite{Spin}. For now, we do not include such
corrections in order to keep our equations reasonably short, noting that they
can always be absorbed in the lapse function at the expense of further
modifying $\kappa(Q)$ and $g(Q)$. Appendix~\ref{a:lapse} demonstrates that
a modified lapse function that takes into account inverse-$a$ corrections in
the $P$-dependent term would not change our main results.

Canonical equations generated by $C$ determine how $P$ is related to
$\dot{Q}=\{Q,NC\}$ for a given lapse function $N$. Upon evaluating this
relationship, inverting it for $P$ as a function of $\dot{Q}$, and inserting
this expression in $C$, we obtain the modified
Friedmann equation \cite{AmbigConstr}
\begin{equation} \label{Friedmann}
 \left(\frac{\dot{a}}{Na}\right)^2 = \frac{8\pi G}{3}
 \left(\frac{m(a)}{\ell_0^3a^3}g(a)- 
   \frac{3}{8\pi G} \frac{\kappa(a)}{\ell_0^2a^2}\right) \left(1+\delta^2
   \frac{a^{4x}}{\ell_0^2}\kappa(a)- \frac{m(a)g(a)}{\ell_0^3a^3 \rho_{\rm
       QG}(a)}\right)\,, 
\end{equation}
transformed back from $Q$ to $a$, with a density scale
\begin{equation} \label{rhoQG}
 \rho_{\rm QG}(a) = \frac{3}{8\pi G \delta^2 a^{2(2x+1)}}
\end{equation}
related to $\delta$.

There is a single constraint in homogeneous models. Therefore, Hamilton's
equations of motion $\dot{Q}=\{Q,NC\}$ and $\dot{P}=\{P,NC\}$ automatically
preserve the constraint, for any lapse function $N$:
$\dot{C}=\{C,NC\}=\{C,N\}C\approx 0$ vanishes when the constraint is
satisfied.  As a consequence, the Friedmann equation of a homogeneous model
can easily be modified, as in (\ref{Cons}), and then automatically generates
consistent continuity and Raychaudhuri equations. However, it is not
guaranteed that these evolution equations correspond to consistent evolution
of an isotropic space-time set up as a background for a covariant theory of
cosmological perturbations. (See \cite{NonCovDressed,Claims} for a discussion
of some subtleties in this context.) For isotropic models, it is sometimes
possible \cite{ActionRhoSquared,LimCurvLQC,HigherDerivLQC} to construct analog
actions which are covariant (of higher-curvature type) and produce modified
Friedmann equations of the form (\ref{Friedmann}). However, these actions fail
to describe holonomy-modified equations of motion in anisotropic models
\cite{MimeticLQC} or for perturbative inhomogeneity \cite{MimeticLQCPert}.

\subsubsection{Perturbation equations and covariance}

The possibility of covariant perturbations on a modified background dynamics
is therefore to be shown and cannot simply be assumed. In a canonical approach
such as loop quantum gravity or its cosmological models, covariance can be
tested systematically by an evaluation of Poisson brackets of constraints for
perturbative inhomogeneity. Because inhomogeneous fields are subject to
multiple constraints, consistency of their equations is not guaranteed: The
Hamiltonian constraint $H[N(x)]$ and diffeomorphism constraint $D[M^a(x)]$
must be such that their Poisson brackets, which generically are not
identically zero and even contain structure functions, vanish when the
constraints are imposed. Moreover, for general covariance to be realized in
the classical or low-curvature limit, their brackets must equal those of
hypersurface deformations \cite{DiracHamGR,ADM},
\begin{eqnarray}
 \{D[M_1^a],D[M_2^b]\} &=& D[{\cal L}_{M_1}M_2^c]\\
 \{H[N],D[M^a]\} &=& -H[{\cal L}_{M_2}N]\\
 \{H[N_1],H[N_2]\} &=& D[\beta q^{ab} (N_1\partial_bN_2-N_2\partial_bN_1)]\,,
\end{eqnarray}
in such a limit, where $q^{ab}$ is the inverse spatial metric and $\beta=\pm
1$ determines space-time signature.  If these brackets are closed after
modifying the constraints, covariance remains intact but as a symmetry it may
receive quantum corrections for instance in the structure function $\beta
q^{ab}$.

Most modifications of holonomy form violate covariance \cite{Disfig}, which
can intuitively be seen from the fact that they only lead to corrections in
terms of powers of $\dot{a}$, while covariant higher-curvature actions would
also require higher time derivatives of $a$ (or auxiliary fields).  For
specific modifications of the terms in an inhomogeneous constraint, it is
possible to respect the closure condition of constraint brackets. However, the
classical brackets are modified by a certain function
\begin{equation} \label{beta}
 \beta(P)=\cos(2\delta P)
\end{equation}
multiplying the bracket of
two Hamiltonian constraints \cite{JR,ScalarHolInv,DeformedCosmo}, assuming
that the background Friedmann equation is modified according to
(\ref{Cons}). The modified theory therefore does not have gauge
transformations of the classical form, which are equivalent to coordinate
changes. Therefore, the form of covariance realized is not one of standard
coordinate changes, at least not for the original metric
variables. Nevertheless, under certain conditions it is possible to apply a
field redefinition of the metric, such that the isotropic line element is not
the usual canonical one,
\begin{equation} \label{ds}
 {\rm d}s^2 = -N^2{\rm d}t^2+ a(t)^2{\rm d}\Omega_k\,,
\end{equation}
but rather \cite{Normal,EffLine}
\begin{equation}
 {\rm d}s_{\beta}^2 = -\beta N^2{\rm d}t^2+ a(t)^2{\rm d}\Omega_k\,.
\end{equation}
When $\beta(P)$ is negative, the line element is positive definite, showing
dynamical signature change. (In full generality, implications of these
modified geometries for space-time structure are still being analyzed
\cite{DefSchwarzschild,DefSchwarzschild2,DefGenBH}.) A modified constraint
(\ref{Cons}), together with trigonometric identities, implies that
\begin{eqnarray}
  \beta&=&\cos(2\delta P)= 1-2\sin^2(\delta P)\nonumber\\
&=& 1-
  2\delta^2\bar{Q}^{-(1-4x)/(2(1-x))}\left(\frac{8\pi G}{3} m(Q)
    g(Q)     -\bar{Q}^{1/(2(1-x))} \kappa(Q)\right)\nonumber\\
&=& 1-2\delta^2 a^{2(2x+1)} \left(\frac{8\pi G}{3}  \rho
  g-\frac{\kappa}{a^2}\right) 
= 1-2\frac{\rho g}{\rho_{\rm QG}(a)}+\frac{3}{4\pi G}
\frac{\kappa}{a^2\rho_{\rm QG}(a)} \label{beta1}
\end{eqnarray}
where $\rho_{\rm QG}(a)$ has been defined in (\ref{rhoQG}).  If $\kappa$ and
$g$ are such that the curvature term $\kappa/a^2$ is negligible compared with
$\rho g$ at small $a$, we have $\beta<0$ for $\rho g>\frac{1}{2}\rho_{\rm
  QG}(a)$. The usual choices of $\delta$ then require Planckian energy
densities for signature change to be realized.

\subsection{Off-shell instantons}
The derivation of (\ref{beta1}) requires an application of the Hamiltonian
constraint (\ref{Cons}) and is not available for off-shell instantons. As one
of the main results reported in \cite{LoopsRescue}, it is nevertheless
possible to derive $\beta$ as a function of $\ddot{Q}$ instead of $P$. Unlike
$P$, $\ddot{Q}$ is available for off-shell instantons because it is determined
by the Raychaudhuri equation without reference to the Friedmann
equation. Signature change is therefore a well-defined space-time phenomenon
even for off-shell instantons, even though they do not satisfy all the
equations implied by generators of hypersurface deformations. In the present
paper, we demonstrate the non-trivial nature of this result.

However, we should first demonstrate the existence of off-shell instantons
with no-boundary initial conditions, $q(0)=0$ and $q(1)=q_1$, after a
modification of background equations.

\subsubsection{Existence}
\label{s:Existence}
Off-shell Lorentzian instantons are determined by solving the second-order
Rauchaudhuri equation for the scale factor $a$ or $q(=a^2)$ without imposing the
first-order Friedmann equation. The usual matter choice in this context is a
cosmological constant, $8\pi Gm/3\ell_0^3=\Lambda a^3=\Lambda q^{3/2}$. Since
there is no inverse of $a$ in such a matter term, $g(a)=1$. Moreover, we
follow \cite{NoRescue} and choose the lapse function $N=M/a=M/\sqrt{q}$.  We
do not directly solve the modified Friedmann equation (\ref{Friedmann}), or
\begin{equation} \label{q1}
 \dot{q}^2=-4M^2\left(\frac{\kappa(q)}{\ell_0^2}-\frac{1}{3}\Lambda q\right)
 \left(1+\delta^2q^{2x} \left(\frac{\kappa(q)}{\ell_0^2}-\frac{1}{3}\Lambda
     q\right)\right) 
\end{equation}
translated to $q$, but first take a second time derivative to obtain
\begin{eqnarray} \label{qdd}
 \ddot{q}&=&\frac{2}{3}\Lambda M^2 \left(1-\frac{3}{\Lambda\ell_0^2} \frac{{\rm
       d}\kappa}{{\rm d}q}\right)
 \left(1+\delta^2q^{2x}\left(\frac{\kappa(q)}{\ell_0^2}-\frac{1}{3}\Lambda
     q\right)\right)\\ 
&&- 4x\delta^2M^2q^{2x-1} \left(\frac{\kappa(q)}{\ell_0^2}-\frac{1}{3}\Lambda
  q\right) 
 \left(\frac{\kappa(q)}{\ell_0^2}
-\frac{1}{6x}\Lambda\left(1+2x-\frac{3}{\Lambda\ell_0^2}
     \frac{{\rm 
         d}\kappa}{{\rm d}q}\right)q\right)\,.\nonumber
\end{eqnarray}

The right-hand side of (\ref{qdd}) contains factors such as $q^{-1}$ ($x=0$) or
$q^{-2}$ ($x=-1/2$), but the full expression is nevertheless regular at $q=0$
if inverse-$q$ corrections are taken into account in $\kappa(q)$. The
no-boundary initial value $q(0)=0$ can therefore be imposed. For small $t$,
$q$ is small and the right-hand side of (\ref{qdd}) is approximately
constant. A generic small-$t$ behavior of $q(t)\propto t+O(t^2)$ then follows.

It is not easy to find exact solutions of (\ref{qdd}) for generic $x$ and
$\delta$.  As a simpler example, we may consider only inverse-triad
corrections ($\delta=0$), in which case the equation reads
\begin{equation}
 \ddot{q}=\frac{2}{3}\Lambda M^2 \left(1-\frac{3}{\Lambda\ell_0^2} \frac{{\rm
       d}\kappa}{{\rm d}q}\right)\,.
\end{equation}
For large $q$, $\kappa\approx \ell_0^2$ and we recover the classical
equation. For small $q$, $\kappa(q)\propto \ell_0^2 q^n$ is an integer power
law in $q$, subject to quantization ambiguities. A simple analytical solution
for $q(t)$ can be found if $\kappa(q)=\kappa_0\ell_0^2 q^2$ is quadratic,
which implies
\begin{equation}\label{qSoln}
 q(t)=\left(q_1+\frac{\Lambda}{6}
   \left(\cos(2\sqrt{\kappa_0}M)-1\right)\right) \frac{\sin(2\sqrt{\kappa_0}
   Mt)}{\sin(2\sqrt{\kappa_0}M)}+
 \frac{\Lambda}{6}\left(1-\cos(2\sqrt{\kappa_0}Mt)\right)
\end{equation}
for no-boundary conditions $q(0)=0$ and $q(1)=q_1$ as in \cite{NoRescue}. A
discussion of stability requires only the small-$t$ behavior,
\begin{equation} \label{q}
 q(t)\approx \frac{2M\sqrt{\kappa_0}}{\sin(2\sqrt{\kappa_0}M)}
 \left(q_1+ \frac{\Lambda}{6}\left(\cos(2\sqrt{\kappa_0}M)-1\right)\right)t
 +O(t^2)\,. 
\end{equation}

As another example, holonomy modifications with $x=-1/2$ can be included if we
assume an inverse-triad correction of the form $\kappa = \kappa_0\ell_0^2
q$. The equation of motion
\begin{eqnarray}
  \ddot{q} = 2 M^2 \frac{\Lambda}{3} \left(1-\frac{3\kappa_0}{\Lambda}\right)
  \left(1 
    + \delta^2\left(\kappa_0- \frac{\Lambda}{3}\right)\right)  
\end{eqnarray}
then gives us a constant $\ddot{q}$, just as in the classical case, and is
solved by
\begin{eqnarray} \label{qsol}
 q(t) &=& t^2 M^2 \left(\frac{\Lambda}{3} - \kappa_0\right) \left(1-\delta^2
   \left(\frac{\Lambda}{3} - \kappa_0\right)\right)\\
&& + t \left(q_1 - M^2
   \left(\frac{\Lambda}{3} - \kappa_0\right) \left(1-\delta^2
     \left(\frac{\Lambda}{3} 
       - \kappa_0\right)\right)\right)\,.\nonumber
\end{eqnarray}

\subsubsection{Off-shell space-time structure}

The analysis of hypersurface-deformation brackets in \cite{ScalarHolInv}
determines $\beta$ through the canonical momentum $P$. As indicated in
(\ref{beta1}), this expression can be written as a function of the energy
density $\rho$ upon using the modified Friedmann equation, but the latter is
not available for off-shell instantons. We will now demonstrate that it is
nevertheless possible to obtain a unique expression for $\beta$, based on the
canonical version of the second-order equation for $q$. In order to capture
potential quantization ambiguities based on the choice of basic variables that
appear in holonomies, we work with the general expressions (\ref{QP}) for $Q$
and $P$. The parameter $x$ in $P$ then determines the $a$-dependence of
holonomy modifications. After $P$ has been eliminated by inserting equations
of motion, we will transform to the variable $q=a^2$ preferred for a
comparison with the Lorentzian path integral.

We first derive the canonical version of modified equations of motion
generated by (\ref{Cons}), again following the choices of \cite{NoRescue}, in
particular for the lapse function $N=M/a$. In this form, the constraint is
given by
\begin{equation}
 \frac{M}{a}C = -\frac{3M\ell_0}{8\pi G} \left(\left(\frac{8\pi G}{3}(1-x)
     Q\right)^{-2x/(1-x)} \frac{\sin^2(\delta P)}{\delta^2}+ \kappa(Q)\right)+
 M\frac{m(Q)g(Q)}{a(Q)}
\end{equation}
and generates first-order equations of motion
\begin{eqnarray}
 \dot{Q} = \{Q,MC/a\}&=& -\frac{3M\ell_0}{8\pi G} \left(\frac{8\pi G}{3}(1-x)
     Q\right)^{-2x/(1-x)} \frac{\sin(2\delta P)}{\delta} \label{Qdot}\\
 \dot{P} = \{P,MC/a\}&=& \frac{3M\ell_0}{8\pi G} \left(-\frac{16\pi G x}{3}
   \left(\frac{8\pi  G}{3}(1-x) 
     Q\right)^{-(1+x)/(1-x)} \frac{\sin^2(\delta P)}{\delta^2}+ \frac{{\rm
       d}\kappa}{{\rm d}Q}\right)\nonumber\\
&&- M\frac{{\rm d}(mg/a)}{{\rm
     d}Q}\,. \label{Pdot} 
\end{eqnarray}
They imply the second-order equation
\begin{eqnarray}
 \ddot{Q} &=& -\frac{3M\ell_0}{8\pi G} \left(-\frac{16\pi G x}{3}
   \left(\frac{8\pi 
       G}{3}(1-x) 
     Q\right)^{-(1+x)/(1-x)} \dot{Q}\frac{\sin(2\delta
     P)}{\delta}\right.\nonumber\\ 
&& \qquad+
   \left.2\left(\frac{8\pi G}{3}(1-x) 
     Q\right)^{-2x/(1-x)} \cos(2\delta P)\dot{P}\right)\nonumber\\
&=& 2\left(\frac{3M\ell_0}{8\pi G}\right)^2 \left(\frac{8\pi G}{3}(1-x)
     Q\right)^{-2x/(1-x)}\nonumber\\
&& \times\left(\frac{8\pi G x}{3\delta^2}
       \left(\frac{8\pi G}{3}(1-x) 
     Q\right)^{-(1+x)/(1-x)}\!\!\! \left(-\sin^2(2\delta P)+ 2\sin^2(\delta
     P)\cos(2\delta P)\right)\right. \label{Qddsquared}\\
&&\quad- \left.\left(\frac{{\rm d}\kappa}{{\rm d}Q}- \frac{8\pi
       G}{3} \frac{{\rm d}(mg/(\ell_0a))}{{\rm d}Q}\right) \cos(2\delta
   P)\right) \,.\nonumber
\end{eqnarray}

Notice that this second-order equation is independent of (\ref{qdd}) because
it is derived from the phase-space expressions of equations of motion. As a
second-order differential equation, it is not complete because it still
contains $P$. The dynamics of $q$ or $Q$ is therefore determined by
(\ref{qdd}), and only by this equation in a consideration of off-shell
instantons. Equation~(\ref{Qddsquared}) then serves as an independent equation
that can be used to determine $\cos(2\delta P)$, or $\beta$ according to
(\ref{beta}), as a function of $Q(t)$.  

The trigonometric identity
$\cos(2\delta P)=1-2\sin^2(\delta P)$ allows us to simplify this expression to
\begin{eqnarray}
\ddot{Q}&=& 2\left(\frac{3M\ell_0}{8\pi G}\right)^2 \left(\frac{8\pi G}{3}(1-x)
     Q\right)^{-2x/(1-x)} \left(\left(\frac{8\pi G x}{3\delta^2}
       \left(\frac{8\pi G}{3}(1-x) 
     Q\right)^{-(1+x)/(1-x)}\right.\right.\label{Qdd}\\
&&- \left.\left.\left(\frac{{\rm d}\kappa}{{\rm d}Q}- \frac{8\pi
       G}{3} \frac{{\rm d}(mg/(\ell_0a))}{{\rm d}Q}\right)\right) \cos(2\delta
 P)- 
 \frac{8\pi 
     G x}{3\delta^2} \left(\frac{8\pi G}{3}(1-x)
     Q\right)^{-(1+x)/(1-x)}\right)\,.\nonumber
\end{eqnarray}
Importantly, we have eliminated all quadratic terms in $\cos(2\delta P)$ or
$\sin(2\delta P)$.  Therefore, we can uniquely solve the equation for
\begin{eqnarray} \label{cos}
 \cos(2\delta P)&=& \frac{\frac{8\pi G x}{3\delta^2} \left(\frac{8\pi G}{3}(1-x)
     Q\right)^{-(1+x)/(1-x)}+ \frac{1}{2}\ddot{Q} \left(\frac{8\pi
       G}{3M\ell_0}\right)^2 \left(\frac{8\pi G}{3}(1-x)
     Q\right)^{2x/(1-x)}}{\frac{8\pi G x}{3\delta^2} \left(\frac{8\pi G}{3}(1-x)
     Q\right)^{-(1+x)/(1-x)}- {\rm d}\kappa/{\rm d}Q+ \frac{8\pi G}{3}
   {\rm d}(mg/(\ell_0a))/{\rm d}Q}\\
&=& \left\{\begin{array}{cl} \displaystyle\frac{1+ \frac{4\pi
       G\delta^2}{3xM^2\ell_0^2} \left(\frac{8\pi G}{3}(1-x)
     Q\right)^{(1+3x)/(1-x)}\ddot{Q}}{1+\frac{\delta^2}{x}\left(\frac{8\pi
       G}{3}(1-x)  Q\right)^{(1+x)/(1-x)}\left({\rm d}(mg/(\ell_0a))/{\rm
       d}Q-\frac{3}{8\pi G} {\rm d}\kappa/{\rm d}Q\right)}
 & \mbox{ if }x\not=0\\
 \displaystyle\frac{\frac{1}{2}\ddot{Q} \left(\frac{8\pi
       G}{3M\ell_0}\right)^2}{ \frac{8\pi G}{3}
   {\rm d}(mg/(\ell_0a))/{\rm d}Q- {\rm d}\kappa/{\rm d}Q} & \mbox{ if
 }x=0\end{array}\right. \nonumber
\end{eqnarray}
in terms of $\ddot{Q}$, without taking a square root.

Writing $\cos(2\delta P)$ suggests that there is still a $P$, which however is
determined only by first-order equations that are not available for off-shell
instantons. For off-shell instantons, we can instead use (\ref{cos}) as the
only equation left in the system that determines $P$. The resulting value of
$P$ may not be real because the right-hand side of (\ref{cos}) is not
restricted to be between one and $-1$. This possibility of imaginary  $P$ is
the analog of an imaginary ${\rm d}q/{\rm d}t$ implied by the classical
first-order equation (\ref{Friedmann}) at the no-boundary initial time, where
$q=0$. As in this case, a complex $P$ here is not problematic in a discussion
of off-shell instantons.

What is important is that (\ref{cos}) uniquely determines $\cos(2\delta P)$ as
a real function, even if $P$ may not be real. This real function can be taken
as a definition of $\beta$ following (\ref{beta}) derived from the results of
hypersurface-deformation brackets for perturbative inhomogeneity. These
results use only the off-shell Poisson brackets of constraints for
inhomogeneous perturbations and therefore remain available for off-shell
instantons. We recall the important feature seen in the final equation
(\ref{Qdd}), which is linear in $\cos(2\delta P)=\beta$, such that no roots
need be taken that could limit the allowed range of values. (The intermediate
step (\ref{Qddsquared}) demonstrates that this result is quite non-trivial.)
This outcome is crucial for our extension of dynamical signature change in
models of loop quantum cosmology to off-shell instantons.

At this point, it is convenient to apply the inverse transformation of
(\ref{QP}) from $Q$ to
\begin{equation}
 q=a^2= \frac{1}{\ell_0^2}\left(\frac{8\pi G}{3}(1-x) Q\right)^{1/(1-x)}\,.
\end{equation}
Using 
\begin{equation} \label{dQdq}
 \frac{{\rm d}}{{\rm d}Q}= \frac{{\rm d}q}{{\rm d}Q} \frac{{\rm d}}{{\rm d}q}=
 \frac{8\pi G}{3} \ell_0^{2(x-1)} q^x \frac{{\rm d}}{{\rm d}q}
\end{equation}
and
\begin{equation} \label{Qddq}
 \ddot{Q}=\frac{3\ell_0^{2(1-x)}}{8\pi G}
 \left(\frac{\ddot{q}}{q^x}-x\frac{\dot{q}^2}{q^{1+x}}\right)\,,
\end{equation}
the resulting expression for the off-shell $\beta=\cos(2\delta P)$ is
\begin{equation} \label{betax}
 \beta = \frac{1+\frac{1}{2}M^{-2} \delta^2x^{-1} \ell_0^{2(1+2x)}q^{1+2x}
   \left(\ddot{q}-x\dot{q}^2/q\right)}{1+\delta^2x^{-1} \ell_0^{4x} q^{1+2x}
   \left(\frac{8\pi G}{3}{\rm d}(mg/(\ell_0\sqrt{q}))/{\rm d}q- {\rm
         d}\kappa/{\rm d}q\right)}
\end{equation}
for $x\not=0$ and
\begin{equation} \label{beta0}
 \beta= \frac{\ell_0^2}{2M^2} \frac{\ddot{q}}{\frac{8\pi G}{3}{\rm
       d}(mg/(\ell_0\sqrt{q}))/{\rm d}q- {\rm d}\kappa/{\rm d}q}
\end{equation}
for $x=0$. (The expression for $x=0$ is independent of $\delta$ but differs
from the classical value one. The classical limit $\beta\to 1$ can be obtained
from (\ref{cos}) if the limit $\delta\to 0$ is taken before $x\to 0$.)

\subsubsection{Renormalization parameters}

There are two common (but non-unique) choices for $x$: If $x=0$, $Q=3\ell_0^2
a^2/(8\pi G)$ is proportional to the isotropic version of a densitized triad,
and $P=- \ell_0\dot{a}$ is the isotropic component of the connection or
extrinsic curvature.  These tensors are used as basic variables in loop
quantum gravity. At a technical level, this case is therefore preferred in
fundamental constructions. However, it implies a fixed co-moving discreteness
scale $\delta$ in the holonomy used to quantize the Hamiltonian constraint
which can easily grow to macroscopic values as the universe expands: Writing
the argument of holonomies in this case as $\delta P=-\delta
\ell_0\dot{a}=-\delta \ell_0 a H$ shows that it can grow very large during an
inflationary period with nearly constant Hubble parameter $H$. 

Using a homogeneous cosmological model in considerations of long evolution
times, in particular during inflation, means that one is applying an effective
description of a fundamental theory on a vast range of scales. It is not
reasonable to expect that the same effective theory, with constant parameters,
remains valid over the whole range. Parameters that describe the effective
dynamics should rather be adjusted, or renormalized, as the scales change.
The averaging of a fundamental state implicitly described by a homogeneous
minisuperspace model is therefore expected to lead to a running $x$ as well as
$\delta$. The effective power-law exponent $x$ may still describe the
effective evolution in sufficiently short periods of time, but it need not be
equal to zero or remain constant.

While the derivation of a running $x$ from a fundamental discrete theory is
challenging, it is possible to model possible outcomes of cosmic evolution by
a succession of phases with different $x$, such that $x$ is nearly constant in
each phase (much like the energy density is usually assumed to be of power-law
form depending on the dominant matter contribution). If $x<0$,
$P=-\ell_0^{2x+1} a^{2x}\dot{a}$ contains a suppression by the scale factor,
such that the increase of the discreteness scale is slowed down compared with
$x=0$. For $x=-1/2$, we have $P=-H$, and the scale remains
constant if $H$ is constant. This value is therefore preferred from the
perspective of model building {\em if} one assumes that a long period of
cosmic evolution can be described without renormalization \cite{APS}. However,
given the general expectation that renormalization does take place, the
argument cannot be used to show that only the value $x=-1/2$ is possible
\cite{Claims}.

New arguments that do not require long cosmic evolution are therefore
necessary if one tries to restrict possible choices of $x$.  An example is the
observation made in \cite{ScalarHolInv} that the value $x=-1/2$ may be
preferred in constructions of consistent holonomy modifications in the
constraints for perturbative inhomogeneity. We are now ready to derive a new
result of this form based on signature change in off-shell instantons.  We
therefore return to our equations (\ref{betax}) and (\ref{beta0}) for $\beta$
depending on $x$, and recall that $\beta<0$ is of advantage in the no-boundary
proposal because it moves the branch cut of (\ref{gamma1}) from the real axis
to the imaginary axis, eliminating the imaginary part of the action evaluated
on no-boundary instantons.

First looking at the case of $x=0$, equation~(\ref{beta0}) evaluated for
small-$t$ no-boundary solutions such that $q(t)\propto t$, we see that
$\beta=0$. While this value differs significantly from the classical behavior,
it does not imply signature change with $\beta<0$.  For $x=-1/2$, assuming as
usual that $8\pi Gm/3\ell_0^3=\Lambda q^{3/2}$ is determined completely by a
cosmological constant $\Lambda$ and ignoring inverse-triad corrections ($g=1$
and $\kappa/\ell_0^2=1$), we obtain
\begin{equation}
 \beta=\frac{1-(\delta/M)^2
   \left(\ddot{q}+\frac{1}{2}\dot{q}^2/q\right)}{1-2\delta^2 \Lambda}
\end{equation}
from (\ref{betax}).  For sub-Planckian $\Lambda$, the denominator is close to
one. For small-$t$ no-boundary solutions, we have $q(t)\approx ct$ as shown in
Section~\ref{s:Existence}, such that
\begin{equation} \label{betapole}
 \beta\approx 1-\frac{\delta^2c}{2M^2} \frac{1}{t}<0
\end{equation}
is negative as long as $t<\frac{1}{2}\delta^2c/M^2$. Since we will now set out
to demonstrate that this version of signature change is able to rescue the
no-boundary proposal, we have obtained another reason why $x=-1/2$ should be
preferred compared with $x=0$, if only these two choices are considered. More
generally, (\ref{betax}) evaluated on small-$t$ off-shell instantons implies
that
\begin{equation}
 \beta= \frac{1-\frac{1}{2}(M\ell_0)^{-2} \delta^2 (\ell_0^2c)^{2(1+x)}
   t^{2x}}{1+x^{-1} \delta^2 (\ell_0^2c)^{1+2x} \Lambda t^{1+2x}}
\end{equation}
for any $x\not=0$. For $x<-1/2$, both numerator and denominator may be
negative for small $t$. The presence of signature change therefore depends on
relationships between parameters such as $\Lambda$ and $\delta$, and is not as
generic as in (\ref{betapole}). The range $-1/2<x<0$ leads to a qualitative
behavior similar to $x=-1/2$, but the phase of signature change becomes
shorter and shorter as $x$ approaches zero because the pole of $t^{2x}$ in the
numerator of $\beta$ then weakens. The value $x=-1/2$ therefore optimizes the
generic nature and duration of signature change, maximally stabilizing
perturbations around off-shell instantons.

As just mentioned, the function (\ref{betapole}) has a pole at $t=0$, which
will imply a subtlety in our detailed stability analysis given in the
following subsection. When transforming (\ref{Qdd}) for $x=-1/2$ from $Q$ to $q$
using (\ref{dQdq}) and (\ref{Qddq}), we encounter the expression
\begin{equation}
 \ddot{q}= -\frac{1}{2}\frac{\dot{q}^2}{q}+ M^2
  \left(
   \frac{16\pi G}{3}\beta \frac{{\rm d}(q\rho)}{{\rm d}q}
   +\frac{1-\beta}{\delta^2}\right)\,,
\end{equation}
evaluated here for $\kappa=\ell_0^2$ and $g=1$, writing $\rho= m/\ell_0^3a^3$.
We can see that the pole of $\beta$ has a direct relationship with the
no-boundary initial condition, which implies that the left-hand side is zero
while the pole of $-\frac{1}{2}\dot{q}^2/q =-\frac{1}{2}c/t$ in the first term
on the right-hand side must be canceled if the equation holds true. The pole
in $\beta$, which changes the asymptotic form of the mode equation
(\ref{vclassapprox}), is therefore directly implied by the no-boundary initial
condition.

\subsubsection{Stability}

Covariant equations compatible with (\ref{Friedmann}) have been derived in
\cite{ScalarHolInv}, as well as in \cite{JR,HigherSpatial} for spherically
symmetric models with closely related properties \cite{DeformedCosmo}.  For
small $t$ and $q$, we assume, as before, that $g(q)=g_0 q^n$ is a power law
with some integer $n$ and positive $g_0>0$. Tensor perturbations $h(\eta)$ in
conformal time $\eta$ are then subject to \cite{ScalarHolInv}
\begin{equation} \label{h}
 h''+\left(2(1+n)\frac{a'}{a}-\frac{\beta'}{\beta}\right) h'-
 \frac{\beta}{\Sigma}\nabla^2h=0
\end{equation}
with
\begin{equation}
 \Sigma=\frac{1}{g_0} \frac{2n-3}{(1+n)(n-3)}\,.
\end{equation}
Since $n$ is an integer, $\Sigma>0$ unless $n=2$. From now on we will assume
the generic case, $n\not=2$ such that $\Sigma$ is positive. 

Again, we transform to $q=a^2$ instead of $a$ and use a time coordinate
according to $N=M/a$, obtaining 
\begin{equation} \label{v}
 \ddot{v}-\left(\frac{\ddot{z}}{z}
   +\frac{\dot{\beta}^2}{4\beta^2}+\frac{\ddot{\beta}}{2\beta}  \right)
   v +\frac{M^2}{q^2} \frac{\beta}{\Sigma} \nabla^2v=0 
\end{equation}
where $z=q^{1+n/2}/\sqrt{|\beta|}$ and $v=zh$.
We first demonstrate how dynamical signature change can lead to stability by
assuming that $\beta$ is nearly constant and negative for some range of small
$q$, which leads to an easy comparison with the results of \cite{NoRescue}.
For a tensor mode of multipole moment $\ell$ and an off-shell instanton
(\ref{q}), the small-$t$ mode equation is then
\begin{eqnarray} \label{vdd}
 \ddot{v}_{\ell}&\approx& \left(\frac{n(n+2)}{4}- \beta \frac{\ell(\ell+2)
     \sin^2(2\sqrt{\kappa_0}M)}{4\Sigma\kappa_0\left(q_1+\frac{1}{6}\Lambda
       \left(\cos(2\sqrt{\kappa_0}M)-1\right)^2\right)}\right)
 \frac{v_{\ell}}{t^2}\\ 
&=& \frac{\gamma_{\ell}^2-1}{4} \frac{v_{\ell}}{t^2}\,. \nonumber
\end{eqnarray}
As in \cite{NoRescue}, there are two independent solutions $v_{\ell,\pm} =
t^{\frac{1}{2}(1\pm\gamma_{\ell})}v_1$, but $\gamma_{\ell}$ is modified. As an
example, we can ignore inverse-$a$ corrections by setting $n=0$ and
$2\sqrt{\kappa_0}M\ll 1$ in (\ref{q}), and obtain
\begin{equation} \label{gamma}
 \gamma_{\ell}=\sqrt{1-4\beta\frac{\ell(\ell+2)M^2}{\Sigma q_1^2}}\,.
\end{equation}
The case of $\beta=1$, used in \cite{NoRescue}, implies branch cuts on the
real $M$-axis for both solutions $v_{\ell,\pm}$. The action for modes
evaluated in these solutions, (\ref{S}), then acquires imaginary parts that
lead to instability.  With effects from loop quantum cosmology, in particular
dynamical signature change, we have $\beta<0$ such that $\gamma_{\ell}$ does
not have branch cuts on the real $M$-axis. Therefore, one does not expect
unstable Gaussians to result from a path integration over $M$.

The full equation (\ref{vdd}) also shows that inverse-$a$ corrections can
change the positions of branch cuts. Even if $\beta>0$, inverse-$a$
corrections can partially improve stability, but real $\gamma_{\ell}$ are then
obtained only for a finite number of multipoles, with a maximum value related
to the ambiguity parameter $n$. If we had only inverse-$a$ corrections, models
of loop quantum gravity would not lead to complete stability. Nevertheless,
including inverse-$a$ corrections in (\ref{vdd}) is useful because it shows
that they do not interfere with stability as implied by holonomy modifications
when $\beta$ is negative.

In the asymptotic regime of very small $t$, which is most relevant for
no-boundary initial conditions, it is not possible to assume nearly constant
$\beta$ because (\ref{betapole}) has a pole at $t=0$. This pole, rather than
the classical $1/t^2$-behavior, dominates the mode equation
\begin{equation} \label{vddalpha}
 \ddot{v}_{\ell}=\left(\frac{\ddot{z}}{z}
   +\frac{\dot{\beta}^2}{4\beta^2}+\frac{\ddot{\beta}}{2\beta}  \right)
 v_{\ell} + \frac{M^2\ell(\ell+2)}{q^2} \frac{\beta}{\Sigma} v_{\ell}\approx 
 \frac{\alpha_{\ell} v_{\ell}}{t^3}
\end{equation}
where $\alpha_{\ell}=\frac{1}{2}\delta^2\ell(\ell+2)/(\Sigma c)>0$.  This
equation can be solved by modified Bessel functions of the second kind, and we
obtain
\begin{equation}  \label{valpha}
 v_{\ell}(t)= \sqrt{t}
 \frac{K_1(\sqrt{\alpha_{\ell}/t})}{K_1(\sqrt{\alpha_{\ell}})} v_1 
\end{equation}
for the regular solution.

The action is now given by
\begin{equation}
 S_{\ell}=\frac{1}{16\pi
   GM}\int_0^1\left(\dot{v}_{\ell}^2+\frac{\alpha_{\ell}}{4t^3}v_{\ell}^2\right){\rm 
   d}t\,. 
\end{equation}
It is convenient to integrate by parts,
\begin{equation}
 S_{\ell}=\frac{1}{16\pi GM}\left.\left(v_{\ell}\dot{v}_{\ell}\right)\right|_{t=0}^1-
 \frac{1}{16\pi GM}
 \int_0^1\left(v_{\ell}\ddot{v}_{\ell}-
   \frac{\alpha_{\ell}}{4t^3}v_{\ell}^2\right){\rm d}t\,, 
\end{equation}
in which the last integral vanishes thanks to the mode equation. We can then
simply insert the regular solution (\ref{valpha}) in the boundary term and use
$K_1'(z)=-K_0(z)-K_1(z)/z$:
\begin{equation}
 S_{\ell}= \frac{1}{32\pi G}
 \frac{\sqrt{\alpha_{\ell}}}{K_1(\sqrt{\alpha_{\ell}})^2} 
 \left.\frac{K_1(\sqrt{\alpha_{\ell}/t})
K_0(\sqrt{\alpha_{\ell}/t})}{\sqrt{t}}\right|_{t=0}^1 
 v_1^2\,. 
\end{equation}
The asymptotic behavior $K_j(z)\sim \sqrt{\pi/2z}\: e^{-z}$ for $z\gg j$ shows
that the action
\begin{equation} \label{Seval}
 S_{\ell}\sim \frac{\pi}{32\pi G K_1(\sqrt{\alpha_{\ell}})^2}
 \left.\exp(-2\sqrt{\alpha_{\ell}/t})\right|_{t=0}^1 v_1^2= \frac{\pi}{32\pi G
   K_1(\sqrt{\alpha_{\ell}})^2} v_1^2
\end{equation}
is finite.

The function (\ref{betapole}) implies signature change if and only if
$\alpha_{\ell}>0$ for all $\ell$. The same condition results in a real action
(\ref{Seval}) without any imaginary part that could lead to instabilities, as
in (\ref{S}). Moreover, the leading asymptotic order in (\ref{vddalpha}) is
completely independent of $M$ because the $M$-dependence of the spatial
derivative term in the general mode equation cancels out with the
$M$-dependence in (\ref{betapole}). The action, therefore, does not have any
branch cuts in the complex $M$-plane.

\section{Conclusions}

Our detailed derivations of mode equations in the Lorentzian no-boundary
proposal for loop quantum cosmology have revealed several subtle features
which conspire to stabilize perturbative inhomogeneity around off-shell
instantons with no-boundary conditions.\footnote{Without incorporating the
  crucial input of dynamical signature-change, one would not be able to see
  these features as has been noted in \cite{LQGNB,LQGNB2}. These studies
  rather focused on providing new insight on how the no-boundary wave function
  can reveal new interesting dynamics in models of loop quantum cosmology by
  setting up novel initial conditions.} In particular, the possibility of
sub-Planckian signature change in off-shell instantons is surprising and
constitutes a new physical effect even though it is based on the same
constraint analysis \cite{ScalarHolInv} as in the standard on-shell treatment
in loop quantum cosmology. The precise form of the signature function
(\ref{betapole}) then showed several important features --- related to its
pole, the sign of its coefficients, and the dependence on the lapse function
--- that played important roles in our stabilization results. The constructive
interplay between loop quantum cosmology and the no-boundary proposal (or any
initial-value formulation in the Lorentzian path integral) is therefore highly
non-trivial.

\section*{Acknowledgements}

This work was supported in part by NSF grants PHY-1607414 and PHY-1912168. SB
is supported in part by the NSERC (funding reference \# CITA 490888-16)
through a CITA National Fellowship and by a McGill Space Institute fellowship.

\begin{appendix}

\section{Inverse-$a$ corrections, absorbed in the lapse function}
\label{a:lapse}

If the $P$-dependent term in the Hamiltonian constraint (\ref{Cons}) contains
inverse-$a$ corrections $\eta(Q)$, they can be absorbed in the lapse
function provided $\kappa(Q)$ and $g(Q)$ are changed accordingly. For most of
our calculations, we worked with general expressions for the latter two
functions, but we assumed that $N=M/\sqrt{q}$ with constant $M$. If
inverse-$a$ corrections are absorbed in the lapse function, our equations will
receive additional terms because the $q$-dependence of $N$ changes. Here, we
show that the resulting equations do not endanger our main result.

In the constraint and Friedmann equations, we can in this case simply replace
$N$ with $N\eta(Q)$, $\kappa(Q)$ with $\kappa(Q)/\eta(Q)$, and $g(Q)$ with
$g(Q)/\eta(Q)$. Equation~(\ref{Friedmann}), for instance, will be
multiplied by $\eta^2$ on the right, and (\ref{beta1}) remains unchanged
because it is obtained by setting the constraint equal to zero, such that any
$N$ or $N\eta$ cancel out.

However, new terms arise as soon as we start taking time derivatives of our
initial equations. While (\ref{q1}) is just modified by using $M\eta$
instead of $M$, the term
\begin{equation}
 -4M^2\eta \frac{{\rm d}\eta}{{\rm d}q}
 \left(\frac{\kappa}{\ell_0^2}-\frac{1}{3}\Lambda 
   q\right) \left(1+\delta^2q^{2x}
   \left(\frac{\kappa}{\ell_0^2}-\frac{1}{3}\Lambda 
     q\right)\right)
\end{equation}
must be added to (\ref{qdd}). Therefore,
\begin{eqnarray}
  \ddot{q} &=& \frac{2}{3} M^2\eta^2\Lambda \left(1-\frac{3}{\Lambda}+
    2\frac{{\rm d}\log\eta}{{\rm d}\log q}- \frac{6\kappa}{\Lambda\ell_0^2} 
    \frac{{\rm d}\log\eta}{{\rm d}q}\right) 
 \left(1+\delta^2q^{2x}\left(\frac{\kappa(q)}{\ell_0^2}-\frac{1}{3}\Lambda
     q\right)\right)\nonumber \\
&&- 4x\delta^2M^2\eta^2q^{2x-1}
\left(\frac{\kappa(q)}{\ell_0^2}-\frac{1}{3}\Lambda q\right) 
 \left(\frac{\kappa(q)}{\ell_0^2}-
\frac{1}{6x}\Lambda\left(1+2x-\frac{3}{\Lambda}     \frac{{\rm 
         d}\kappa}{{\rm d}q}\right)q\right)\,.
\end{eqnarray}

In (\ref{Qdot}), we again simply replace $M$ with
$M\eta$. In (\ref{Pdot}), we make the same replacement, but also add the
term
\begin{equation}
 \frac{3M\ell_0}{8\pi G}\frac{{\rm d}\eta}{{\rm d}q} \left(\left(\frac{8\pi
       G}{3}(1-x)Q\right)^{-2x/(1-x)} \frac{\sin^2(\delta P)}{\delta^2}+
   \kappa\right)- M \frac{mg}{\sqrt{Q}}
\end{equation}
implied by a $Q$-derivative of the constraint.
The additional time derivative taken to derive (\ref{Qddsquared}) leads to
further terms, such that now
\begin{eqnarray}
\ddot{Q} &=& 2\left(\frac{3M\ell_0\eta}{8\pi G}\right)^2 \left(\frac{8\pi
    G}{3}(1-x) 
     Q\right)^{-2x/(1-x)}\nonumber\\
&& \times\left(\frac{8\pi G x}{3\delta^2}
       \left(\frac{8\pi G}{3}(1-x) 
     Q\right)^{-(1+x)/(1-x)} \left(-\sin^2(2\delta P)+ 2\sin^2(\delta
     P)\cos(2\delta P)\right)\right.\\
&&\quad- \left(\frac{{\rm d}\kappa}{{\rm d}Q}- \frac{8\pi
       G}{3} \frac{{\rm d}(mg/\ell_0a)}{{\rm d}Q}\right) \cos(2\delta
   P) \nonumber\\
&&\quad+ \frac{{\rm d}\log \eta}{{\rm d}Q} \left(\left(\frac{8\pi
        G}{3}(1-x) 
     Q\right)^{-2x/(1-x)} \frac{1}{2\delta^2}\left(\sin^2(2\delta P)-
     2\sin^2(\delta P)\cos(2\delta P)\right)\right.\nonumber\\
 &&\quad+ \left.\left.\kappa-\frac{8\pi
     G}{3}\frac{mg}{\ell_0a(Q)} \right)\right)\,.\nonumber
\end{eqnarray}
The same trigonometric identities as in (\ref{Qddsquared}) then imply that
$\ddot{Q}$ depends linearly on $\cos(2\delta P)$, and our remaining results go
through. 

\section{Saddle-point analysis}

The background on-shell action for the solution (\ref{qsol}), here setting
$\kappa_0=1$, has a strikingly similar form compared with the classical one,
\begin{eqnarray}
	S_0 &=& -\frac{3 q_1^2}{4 M} +  \frac{1}{2} M q_1 \left(-3 + \delta^2
          \left(\Lambda -3\right)\right) \left(\frac{\Lambda}{3} -
          1\right)\nonumber\\
&&  +
        \frac{1}{324} M^3 (\Lambda -3)^2  \left(\delta ^2 (\Lambda
          -9)-3\right) \left(\delta ^2 (\Lambda +3)-3\right)\,. 
\end{eqnarray}
For saddle points, $\partial S_0/\partial M = 0$, we once again get four
solutions. Their analytic forms are still somewhat involved,
\begin{eqnarray}
  M &=& -\frac{3 \sqrt{q_1}}{\sqrt{(3-\Lambda) \left(\delta ^2 (\Lambda
        +3)-3\right)}},\\ 
  M &=& \frac{3 \sqrt{q_1}}{\sqrt{(3-\Lambda) \left(\delta ^2 (\Lambda
        +3)-3\right)}},\\ 
  M &=& -\frac{3 \sqrt{q_1}}{\sqrt{(3-\Lambda) \left(\delta ^2 (\Lambda
        -9)-3\right)}},\\ 
  M &=& \frac{3 \sqrt{q_1}}{\sqrt{(3-\Lambda) \left(\delta ^2 (\Lambda
        -9)-3\right)}}\,. 
\end{eqnarray}
On analyzing the conditions for their denominators to
remain real, we find that
\begin{equation}
  |\delta| <\frac{1}{\sqrt{\Lambda/3+1}} \quad \mbox{if } \Lambda>3
\end{equation}
or
\begin{equation} \label{Sol2}
|\delta| >\frac{1}{\sqrt{\Lambda/3+1}} \quad \mbox{if } \Lambda<3
\end{equation}
for the first two solutions, and
\begin{equation}
  |\delta| <\frac{1}{\sqrt{\Lambda/3-3}} \quad \mbox{if } \Lambda>9
\end{equation}
or
\begin{equation}
  |\delta| >\frac{1}{\sqrt{\Lambda/3-3}} \quad \mbox{if } 3<\Lambda<9
\end{equation}
for the other two, while $\Lambda<3$ does not imply real solutions in this
case. 

Thus, for sub-Planckian $\Lambda$, there are at least two imaginary
solutions, and all four solutions are imaginary if $|\delta|<1/\sqrt{2}$ (the
limiting case for $\Lambda\to3$ in (\ref{Sol2})). For larger $\Lambda$, all
four solutions may be real provided $|\delta|<1/\sqrt{2}$.

Without holonomy modification, $\delta=0$, there are only two saddle-point
solutions, both either real or purely imaginary depending on the value of
$\Lambda$. Interestingly, in none of these cases does the reality of the
saddle points depend on the value of $q_1$, unlike in Einstein gravity.  The
reason for this is that we assume $\kappa/\ell_0^2 =q$ for the spatial
curvature in our specific solution, which is a possible behavior of the
inverse-triad term only near $q\sim 0$ (and assuming a specific power-law
behavior). As we go to larger $q$, especially near $q=q_1$, $\kappa \approx
1$, as it should, and analytical solutions are more difficult to come by.

\end{appendix} 

%\bibliographystyle{preprint}
%\bibliography{Bib/QuantGra}

\end{document}